\documentstyle[twocolumn,prl,aps,floats]{revtex}
\input psfig

\def\bea{\begin{eqnarray}}
\def\eea{\end{eqnarray}}
\def\ben{\begin{equation}}
\def\een{\end{equation}}
\def\benu{\begin{enumerate}}
\def\enu{\end{enumerate}}

\def\n{n}

\def\sss{\scriptscriptstyle\rm}

\def\g{_\gamma}

\def\l{^\lambda}
\def\lfc{^{\lambda=1}}

\def\1var{(\bx_1...\bx\N)}

\def\br{{\bf r}}

\def\bx{{x}}

\def\x{_{\sss X}}
\def\c{_{\sss C}}
\def\s{_{\sss S}}
\def\xc{_{\sss XC}}

\def\N{_{\sss N}}

\def\ee{_{\rm ee}}

\def\sph_int{ {\int d^3 r}}

\def\PRA{Phys. Rev. A\ }

\def\JCP{J. Chem. Phys.\ }

\input rotate
\def\derp#1{\frac {d #1 [\n\g]}{d\gamma} \Big|_{\gamma=1}}
\def\der#1{\frac {d #1 [\n\g]}{d\gamma}}
\def\n{\rho}

\begin{document}

\title
{Adiabatic connection from accurate wavefunction calculations}
\author{Derek Frydel and William H. Terilla}
\address{Department of Chemistry, Rutgers University, 315 Penn Street,
Camden, NJ 08102}
\author{Kieron Burke}
\address{Departments of Chemistry and Physics, Rutgers University, 610
Taylor Road, Piscataway, NJ 08854}
\date{Submitted to J. Chem. Phys., Oct. , 1999}
\maketitle
\begin{abstract}
An extremely easy method for accurately calculating the 
adiabatic connection of density functional
theory is presented, and its accuracy tested on both Hooke's
atom and the He atom.   The method is easy because
calculations are needed only for
different values of parameters in the external potential, 
which can be achieved with almost any electronic structure code.
\end{abstract}

\section{Introduction}
\label{intro}

Density functional theory has become a popular
computational method in quantum chemistry, because of its
ability to handle large molecules accurately
but relatively inexpensively\cite{MOAS99,Sb99}.
This success is based on the availability of reliable accurate
approximate functionals, and there is
a constant need for still further accuracy.
The goal of atomization energy errors being reliably less than
1 kcal/mol has not yet been achieved.

An important step forward in this search for
accuracy came when Becke mixed a fraction of
exact exchange with a generalized gradient
approximation (GGA), and reduced errors by a factor
of two or three\cite{Bb93}.
Such hybrid functionals, e.g., B3LYP, are now in common use,
but their underlying justification comes from the 
adiabatic decomposition of density functional
theory.  Initially, the mixing parameters used were 
determined empirically.  Later, it was shown that 
these parameters could be derived non-empirically\cite{BEP97,E96},
and that a single universal mixing coefficient (25\%)
could be rationalized based on the performance of
MP theory for molecules\cite{PEB96}.  Most recently, new functionals
have been proposed which use this adiabatic decomposition
in much detail\cite{SPK99}.

The adiabatic decomposition of an electronic system
is very simple conceptually.  Imagine multiplying the
electron-electron repulsion by a coupling constant $\lambda$.
Now imagine varying $\lambda$, while
keeping the electron density $\n(\br)$ fixed.  This differs
from traditional perturbative methods, 
e.g., M{\'o}ller-Plesset
\cite{MP34},
because
the external potential must be altered at each $\lambda$
to keep the density fixed.  At $\lambda=1$, we have the
physical, interacting electronic system.  But as $\lambda$
is reduced to zero, keeping the density fixed, the 
electrons become those of the non-interacting, Kohn-Sham
system, and
the potential morphs into the Kohn-Sham potential.
All Kohn-Sham DFT calculations are actually performed on
this non-interacting system, and the physical ground-state
energy deduced from it.  The adiabatic connection provides
a continuous connection between the interacting system and its
Kohn-Sham analog.

The only part of the total energy which must be approximated
in such a Kohn-Sham calculation is the exchange-correlation
energy, $E\xc[\n]$, as a functional of the (spin) density.
A further value of the coupling constant is that, through the
Hellmann-Feynman theorem, this energy can be written as an
intergal over the purely potential contribution\cite{LP75,GL76}:
\ben
E\xc= \int_0^1 d\lambda\, U\xc(\lambda),~~~
U\xc(\lambda)=\langle \Psi\l |{\hat V}\ee | \Psi\l \rangle -U,
\label{acm}
\een
where ${\hat V}\ee$ is the Coulomb interaction between
electrons, $\Psi\l$ is the wavefunction at coupling constant
$\lambda$, and $U$ is the Hartree electrostatic energy.
This integral is the adiabatic connection formula.
Hybrid functionals are based on the fact that this integrand,
when applied to the exchange-correlation contribution 
to a bond dissociation energy, is often {\em not} well-approximated by
GGA's, due to their lack of static correlation\cite{BB95,EPB97}.  This
can be partially corrected for most molecules by mixing
in a fraction of exact exchange.  Construction of
functionals based on this insight is often referred
to as the adiabatic connection method (ACM).
\begin{figure}[htb]
\unitlength1cm
\begin{picture}(12,6.5)
\put(-5.2,4.0){\makebox(12,6.5){
\includegraphics{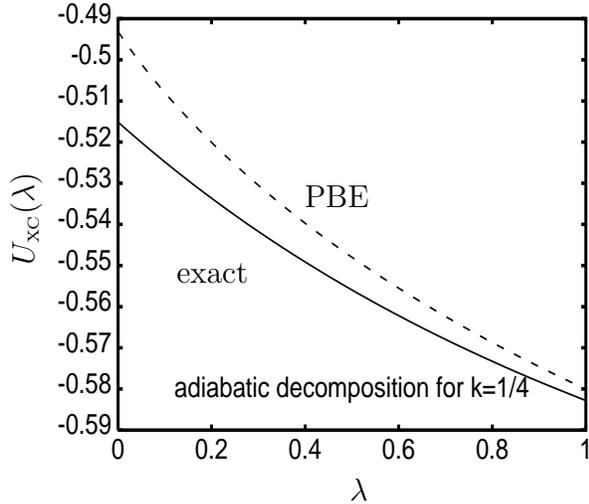}}}
\setbox6=\hbox{\large $U\xc(\lambda)$}
\put(0.8,3.85) {\makebox(0,0){\rotl 6}}
\put(5.1,0.1){\large $\lambda$}
\put(2.8,3.0){\large exact}
\put(4.5,4.0){\large PBE}
\end{picture}
\label{Uxcl}
\caption{Exact $U\xc(\lambda)$ curve for the k=1/4 Hooke's atom,
(solid line) and within the PBE correlation functional (dashed line),
calculated as described in the text (atomic units).}
\end{figure}

Thus accurate approximation of this adiabatic connection curve
is extremely important to further progress in construction
of approximate functionals, and benchmark cases for small
systems are always of interest and help in this endeavor.
However, this is, in principle,
a very demanding task.  For each value of $\lambda$, one must
solve the interacting electronic problem many times
in order to find the external potential which reproduces the
$\lambda=1$ density.  
Almbladh and Pedroza\cite{AP84} made early heroic
attempts, but not with the accuracy of modern calculations.
In the last several years, as the importance of these
curves has become apparent, several groups in different areas
have performed adiabatic decomposition calculations. 
Hood et al.\cite{HCWR98} have calculated
the curve for bulk Si.  Colonna and Savin\cite{CS99} have used the slightly different
Harris and Jones\cite{HJ74} decomposition for both the He and Be isoelectronic
series.  Joubert and Srivastava\cite{JS98} have
used Hylleras-type wavefunctions to
calculate these curves for the  He isoelectronic series.

All these methods require solving the interacting electronic
problem with a potential that differs from the original external
potential, e.g., $-Z/r$ for atoms.  This typically
makes them difficult to transfer to other systems, radically different
from the original one, in which other codes and approximations are being
used.
In this paper,
we show how to construct the adiabatic connection formula
for any atom accurately
by doing calculations simply for different values of $Z$, so that
no modification need be made to an existing wavefunction code.
A standard calculation is simply run for several
different nuclear charges, running
from the physical value up to $Z=\infty$.  
The main focus of this paper is to demonstrate the accuracy of the method.
The solid line in 
Figure 1 is an essentially exact curve for $U\xc (\lambda)$ for 
Hooke's atom (two electrons in a harmonic potential),
for a spring constant of $k=1/4$ (atomic units),
calculated from a series of exact calculations for force
constants greater than 1/4, but never requiring any calculation
with a different external potential.
The method  can be immediately applied to accurate calculations
for larger atoms, and is currently being explored for molecules.

The paper is divided into several sections.  In the next section,
we outline the basic theory behind our calculations.  Following that,
we present results, both using the exact functional, and within
a GGA, for both Hooke's atom and the He atom.  Hooke's atom consists
of two electrons attached by springs to a center, but interacting
via a Coulomb repulsion.    We summarize our findings in the 
last section.
Atomic units ($e^2=\hbar=m_e=1$) are used throughout, and
only spin-unpolarized systems are discussed.

\section{Theory}
\label{theory}

\subsection{Formally exact results}
\label{exact}
For convenience, we begin with a review of the adiabatic connection
formalism.  We define $\Psi\l[\n]$ to be that wavefunction which
has density $\n(\br)$ and minimizes
\ben
F\l[\Psi] = 
\langle \Psi |  {\hat T} +
\lambda {\hat V}\ee | \Psi  \rangle.
\label{Fl}
\een
Then the kinetic-correlation  energy at $\lambda$ is
\ben
T\l\c = \langle \Psi\l | {\hat T} | \Psi\l  \rangle -T\s[\n]
\label{Tlc}
\een
where $T\s[\n]$ is the non-interacting Kohn-Sham kinetic
energy of density $\n$.  The potential contribution to
exchange-correlation at coupling constant $\lambda$ is,
\ben
U\xc\l = \lambda \langle \Psi\l |{\hat V}\ee| \Psi\l  \rangle - \lambda
U[\n],
\label{Uxcldef}
\een
while the  exchange-correlation energy at $\lambda$ is
\ben
E\xc\l = U\xc\l + T\c\l.
\label{Excl}
\een
For $\lambda=1$, the system becomes the true interacting
quantum-mechanical system, and energies without superscripts
refer to $\lambda=1$, e.g., $E\xc=E\xc\lfc$.

All quantities at coupling constants different from one are
simply related to their full coupling strength counterparts,
but evaluated on a scaled density\cite{LP85}.  The most important examples
are the wavefunction:
\ben
\Psi\l [\n] = \Psi_\lambda [\n_{1/\lambda}],
\label{Psil}
\een
and the exchange-correlation energy
\ben
E\xc\l [\n] = \lambda^2 E\xc [\n_{1/\lambda}],
\label{Excscal}
\label{fund}
\een
where $\n\g(\br)=\gamma^3 \n (\gamma \br)$, and 
$\Psi\g (\br_1..\br_N)$ $ =
\gamma^{3/2} \Psi\g (\gamma\br_1..\gamma\br_N)$.
Thus knowledge of how a quantity varies as the density
is scaled implies knowledge of its coupling constant dependence.

Quantities evaluated on the Kohn-Sham wavefunction vary in a
simple way, such as the non-interacting kinetic energy $T\s$:
\ben
T\s\l[\n] = T\s[\n],~~~{\rm or}~~~T\s [\n\g] = \gamma^2\, T\s[\n],
\label{Tsscal}
\een
the Hartree electrostatic energy,
\ben
U\l[\n]=\lambda U[\n],~~~{\rm or}~~~U[\n\g]= \gamma U[\n],
\label{Uscal}
\een
and  the exchange energy
\ben
E\x\l[\n] = \lambda E\x[\n],~~~{\rm or}~~~E\x[\n\g]=\gamma E\x[\n].
\label{Exscal}
\een
Correlation is more sophisticated.  Note, however, that knowledge
of any quantity, $E\c [\n\g], T\c [\n\g]$, or $U\c [\n\g]$
as a function of $\gamma$ for $\gamma$ between 1 and $\infty$,
i.e., scaling to the high density limit, is sufficient to determine
the adiabatic connection for any of them, for $\lambda$ between 0
and 1.  The most well-known relation is to extract the kinetic-correlation
piece from $E\c[\n]$ 
\cite{LP85}:
\ben
T\c[\n] = - E\c[\n] + \derp{E\c} .
\label{TcfromEc}
\een
To generalize this result to $\n\g$,
apply Eq. (\ref{TcfromEc})
to $\n\g$, and make a change of variables in the derivative, to find:
\ben
T\c[\n\g] = - E\c[\n\g] + \gamma \der{E\c}.
\label{TcfromEcg}
\een
This can be considered a first-order differential equation in $\gamma$
for $E\c[\n\g]$.  Solution of this equation yields
\ben
E\c[\n\g] =
- \gamma \int_\gamma^\infty \frac{d\gamma'}{\gamma'^2}\, T\c[\n_{\gamma'}],
\label{EcfromTcg}
\een
where we have used the fact that $E\c[\n\g]$ has vanished as
$\gamma\to 0$
\cite{L91}.
Via Eq. (\ref{fund}), these can be turned into coupling-constant
relations:
\ben
T\c\l = E\c\l - \lambda \frac{d E\c\l}{d\lambda},
\label{TclfromEcl}
\een
and
\ben
E\c\l = - \lambda \int_0^\lambda \frac{d\lambda'}{\lambda'^2}\, T\c^{\lambda'}.
\label{EclfromTcl}
\een
The latter is called Bass' relation\cite{B85}.
Similarly, we can extract $U\c$ from $E\c[\n\g]$, since $U\c=E\c-T\c$.
Thus Eq. (\ref{TcfromEcg}) leads to 
\ben
U\c[\n\g] = 2 E\c[\n\g] - \gamma \der{E\c}
\label{UcfromEcg}
\een
which, inverted, is
\ben
E\c[\n\g]=\gamma^2 \int_\gamma^\infty \frac{d\gamma'}{\gamma'^3}\, U\c[\n_{\gamma'}].
\label{EcfromUcg}
\een
Combining Eqs. (\ref{UcfromEcg}) and (\ref{EcfromTcg})
then leads to simple relations between $U\c$ and $T\c$.
Lastly, the coupling-constant relation that follows from Eq.
(\ref{EcfromUcg}) is 
\ben
E\c[\n] = \int_0^1 \frac{d\lambda}{\lambda} U\c\l[\n],
\label{EcfromUcl}
\een
and $U\xc(\lambda)$ of Fig. 1 is just the integrand when this expression
is applied to both exchange and correlation,
i.e., $U\xc\l/\lambda$.
We emphasize that all these relations follow from the
well-known Eqs. (\ref{fund})
and (\ref{TcfromEc}).

\subsection{Highly accurate approximations}

We begin this section by noting how, when some parameter
in an external potential is altered, the density changes
scale, but often does not change shape very much.  For example,
for the two-electron ion, going from Z=2 to Z=4 will roughly
multiply the density by 8, and reduce its length scale by
a factor of 2.  We use this fact to very accurately approximate
scaling the density.  Then, through the relations derived above,
we can convert this into the coupling-constant dependence.

Let $\n(\br)$ be the density of the system we are interested
in.  Suppose we alter some parameter in the external potential,
and solve the interacting problem, finding some density $\n'(\br)$.
If we want to treat $\n'$ as an approximation to $\n\g$, we
must first choose a criterion for determining $\gamma$.   In fact,
either of Eqs. (\ref{Tsscal}) and (\ref{Uscal}) could be used,
since both would be satisfied
exactly if $\n'$ were truly a scaled density.  For definiteness, we
choose
\ben
\lambda=1/\gamma=U[\n]/U[\n'].
\label{ldef}
\een
We can then consider $E\c[\n'] \approx E\c[\n\g]$, $T\c[\n'] \approx
T\c[\n\g]$, etc., which we call the bare estimates.
As we shall show in the results section,
these energies usually yield quite accurate approximations to the
exact quantities.  However, since we are not working with the true
scaled density, these energies do not satisfy the relations
involving scaling derivatives above, such as Eq. (\ref{EcfromTcg}).

To make a better estimate, we note that, in the case of the
correlation energy, we can calculate the leading correction to
$E\c[\n\g]$, using the correlation potential.  This can be constructed
by finding the exact Kohn-Sham potential which generates the density,
and subtracting from it the external, Hartree,
and exchange contributions.  Then,
\ben
E\c[\n\g] \approx E\c[\n'] + \int d^3r\, v\c[\n'] (\br) \,
\big(\n\g(\br)-\n'(\br)\big) + O(\delta\n)^2.
\label{Eccorr}
\een
As will be shown below, inclusion of this correction leads
to extremely accurate adiabatic connection curves.  We 
generate $T\c[\n\g]$ using Eq. (\ref{EcfromTcg}),
and then $U\c=E\c-T\c$.  In particular, as $\gamma\to 1$,
$\n' \to \n$, and Eq. (\ref{Eccorr}) becomes exact to first order
in the difference between $\n'$ and $\n$.  Thus Eq. (\ref{TcfromEc})
is satisfied exactly.

Lastly, we apply the same principles to $T\s$, as a test of the
closeness of the approximate density, $\n'$.   The Euler equation
for the Kohn-Sham system says that
\ben
\frac{\delta T\s}{\delta \n(\br)} = \mu - v\s(\br),
\label{Tspot}
\een
where $\mu$ is a constant, and $v\s(\br)$ is the 
Kohn-Sham potential.
Thus, we can also correct the bare $T\s$ estimate, to
\ben
T\s[\n\g] = T\s[\n'] - \int d^3r\, v\s [\n'](\br)\, 
\big(\n\g(\br)-\n'(\br)\big) + O(\delta\n)^2.
\label{Tscorr}
\een

\section{Results}
\label{results}

\begin{figure}[htb]
\unitlength1cm
\begin{picture}(12,6.5)
\put(-5.2,4.0){\makebox(12,6.5){
\includegraphics{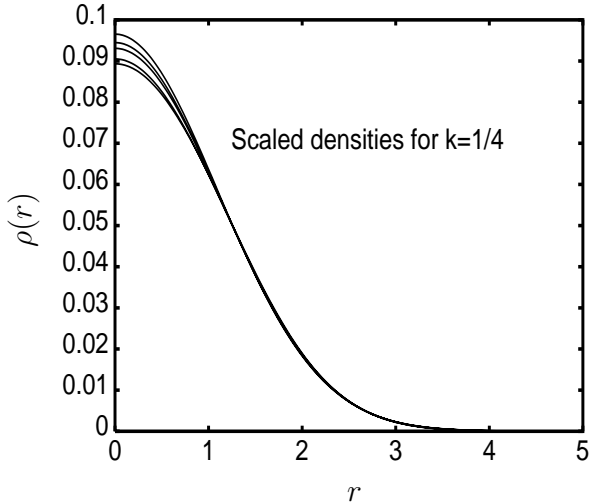}}}
\setbox6=\hbox{\large $\n(r)$}
\put(0.8,3.85) {\makebox(0,0){\rotl 6}}
\put(5.1,0.1){\large $r$}
\end{picture}
\label{dens}
\caption{Densities used for $\lambda=1$ (least $\n(0)$), 
0.75, 0.5, 0.25, and $0$ (largest $\n(0)$).}
\end{figure}
We present results for two systems, Hooke's atom and the He atom.
The first of these contains two electrons in a harmonic potential,
interacting via a Coulomb repulsion.  This provides a valid test
case for density functional theory, because of the Coulomb repulsion
between electrons.  It is also an easy system to perform calculations
on, because the center-of-mass and relative coordinates separate,
leaving only a one-dimension differential equation to solve 
numerically\cite{IBL99}.
Even this equation has an analytic solution for force constant $k=1/4$
\cite{T93},
which is the system on which we will demonstrate most of our results.

To implement our method for the $k=1/4$ Hooke's atom,
we run many calculations at different
values of $k > 1/4$, up to $k=10^6$, using the method of
Ref. \cite{IBL99}.  For each calculation, we
scale the density so that it looks like the original $k=1/4$ density,
using Eq. (\ref{ldef}) to define the scale factor.  Several such densities are
shown in Fig. 2, which illustrates how close these densities are.
Note that the largest error is for $\lambda=0$, the non-interacting
Gaussian density.

\begin{figure}[htb]
\unitlength1cm
\begin{picture}(12,6.5)
\put(-5.2,4.0){\makebox(12,6.5){
\includegraphics{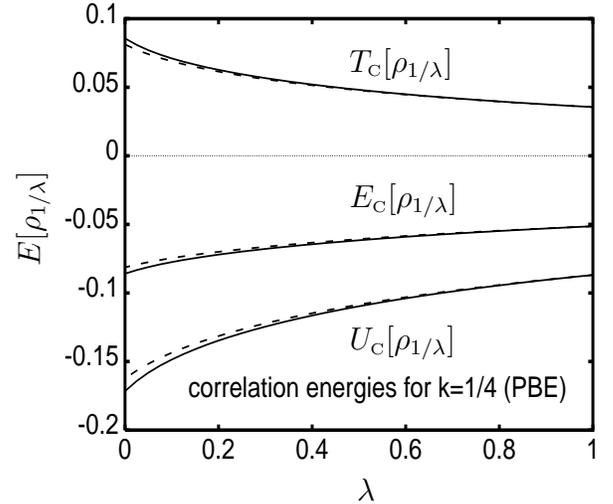}}}
\setbox6=\hbox{\large $E[\n_{1/\lambda}]$}
\put(0.8,3.85) {\makebox(0,0){\rotl 6}}
\put(5.1,0.1){\large $\lambda$}
\put(5.0,4.0){\large $E\c[\n_{1/\lambda}]$}
\put(5.0,2.1){\large $U\c[\n_{1/\lambda}]$}
\put(5.0,5.8){\large $T\c[\n_{1/\lambda}]$}
\end{picture}
\label{ecpbe}
\caption{Correlation energies for the k=1/4 Hooke's atom, calculated
both with (solid lines) and without (dashed lines) the correction 
term, using the PBE correlation functional.  The solid lines are indistinguishable
from exact results, calculated by scaling the PBE correlation functional.}
\end{figure}
Figure 3 is a plot of the three quantities $E\c[\n_{1/\lambda}]$,
$T\c[\n_{1/\lambda}]$, and $U\c[\n_{1/\lambda}]$, all found
using the PBE correlation functional.  We choose these
quantities to plot, as they remain finite in the range $\lambda=0$
to $1$. 
For each quantity, there are two curves.  The dashed curve 
is the bare result of the calculations at different values of
the force constant, i.e., using $E[\n']$ alone.  The solid curve
for $E\c[\n_{1/\gamma}]$ includes
the correction due to the potential, in accordance with Eq. (\ref{Eccorr}).
The $U\c$ and $T\c$ curves are then extracted from this one,
using Eqs. (\ref{TcfromEcg}) and (\ref{UcfromEcg}).
Both the value and first derivative at $\lambda=1$ are
{\em exact} for this curve.
What is more remarkable is that these
curves coincide essentially exactly with the correct result,
which we can deduce by directly applying Eq. (\ref{fund}) to the PBE correlation
energy functional.  Careful numerical calculations indicate that
the maximum error in our curve is at $\lambda=0$, and is less
than 3$\times 10^{-4}$ Hartrees.
This is due to the similarity of densities, as shown in Fig. 2,
leading to very small corrections.

We also tested other possible prescriptions for choosing 
$\lambda$, such as from the square root of the ratio
of non-interating kinetic energy densities, as in Eq. (\ref{Tsscal}).
This gives values for $\lambda$ very close to those of Eq. (\ref{Uscal}),
and leads to no measurable change in our estimate for 
$E\c[\n_{1/\lambda}]$.

\begin{figure}[htb]
\unitlength1cm
\begin{picture}(12,6.5)
\put(-5.2,4.0){\makebox(12,6.5){
\includegraphics{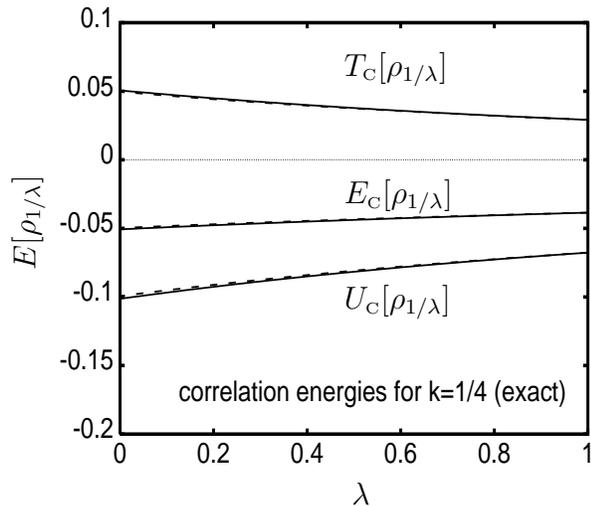}}}
\setbox6=\hbox{\large $E[\n_{1/\lambda}]$}
\put(0.8,3.85) {\makebox(0,0){\rotl 6}}
\put(5.1,0.1){\large $\lambda$}
\put(5.0,4.1){\large $E\c[\n_{1/\lambda}]$}
\put(5.0,2.7){\large $U\c[\n_{1/\lambda}]$}
\put(5.0,5.8){\large $T\c[\n_{1/\lambda}]$}
\end{picture}
\label{ecexa}
\caption{Correlation energies for the k=1/4 Hooke's atom, calculated
both with (solid lines) and without (dashed lines) the correction 
term.}
\end{figure}
Figure 4 is the analog of Fig. 3, but now calculated using
exact correlation energies and potentials.  Based on the remarkable
accuracy of Fig. 3, we claim these curves are essentially exact.
In fact, the change due to the potential correction is much smaller
for the exact case than for PBE,
suggesting that the error made in Eq. (\ref{Eccorr}) will also
be smaller.
The adiabatic curve of Fig. 1 was derived from these curves,
since
$U\xc(\lambda)=\lambda U\xc [\n_{1/\lambda}]$.

We also ran PBE calculations for lower densities, where the
shape of the density can change significantly with $k$.
We find that even as low as $k=10^{-4}$, there is a maximum
error of only 1 millihartree in the corrected $E\c[\n_{1/\lambda}]$
curve.  Beyond this point, the reliability of our method might be
questioned.  However, defining an average $r_s$ value via Eq.
(6a) of 
Ref. \cite{ZBEP97}, 
at this point $\langle r_s \rangle = 19$, whereas
for $k=1/4$, $\langle r_s \rangle$ is 2.07.  Thus for common
values of the density, the errors in our procedure are minute.

\begin{figure}[htb]
\unitlength1cm
\begin{picture}(12,6.5)
\put(-5.2,4.0){\makebox(12,6.5){
\includegraphics{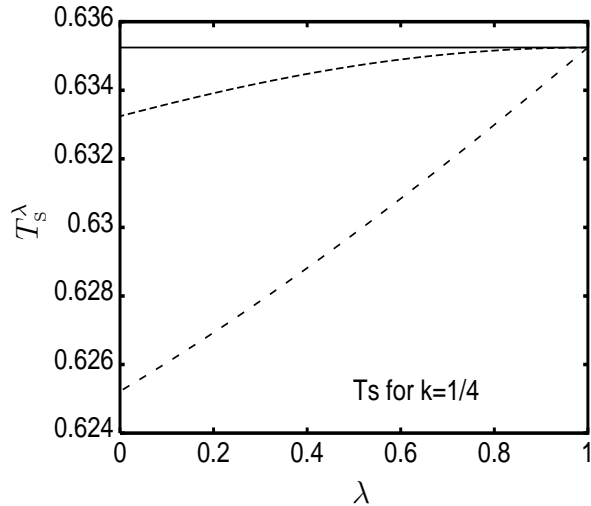}}}
\setbox6=\hbox{\large $T\s\l$}
\put(0.8,3.85) {\makebox(0,0){\rotl 6}}
\put(5.1,0.1){\large $\lambda$}
\end{picture}
\label{Tstest}
\caption{Non-interacting kinetic energy as a function of coupling
constant.  Exact quantity (solid line) is independent of $\lambda$.
Dashed line is bare estimate of $T\s$, the dotted line includes the
correction.}
\end{figure}
As a final test of our method, we return to the non-interacting
kinetic energy.   If the scaling were exact, the 
non-interacting kinetic energy
would scale quadratically, as in Eq. (\ref{Tsscal}). 
In Figure 5, we plot $T\s[\n\g]/\gamma^2$ exactly,
approximating $T\s[\n\g]$ by $T\s[\n']$, and including
the correction of Eq. (\ref{Tscorr}).  Note that the maximum absolute error,
at $\lambda=0$, is only about -1.6\%, in the bare estimate.
The correction makes the derivative with respect to $\lambda$
exact as $\lambda\to 1$, and overall reduces the error
by about a factor of 5, to about -0.3\%.

\begin{figure}[htb]
\unitlength1cm
\begin{picture}(12,6.5)
\put(-5.2,4.0){\makebox(12,6.5){
\includegraphics{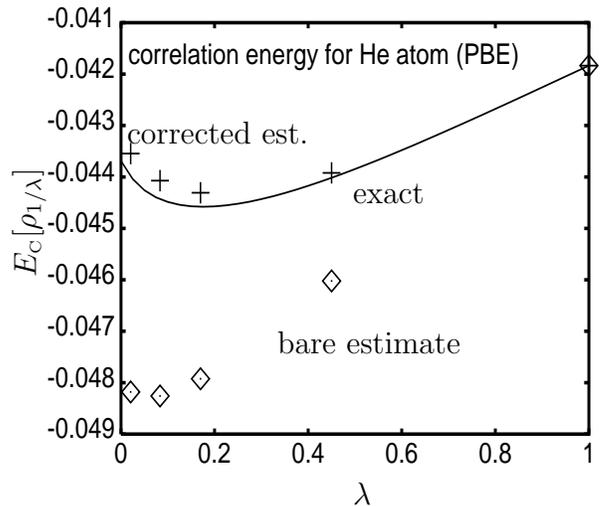}}}
\setbox6=\hbox{\large $E\c[\n_{1/\lambda}]$}
\put(0.8,3.85) {\makebox(0,0){\rotl 6}}
\put(5.1,0.1){\large $\lambda$}
\put(4.1,2.1){\large bare estimate}
\put(2.1,4.9){\large corrected est.}
\put(5.1,4.1){\large exact}
\end{picture}
\label{ecHepbe}
\caption{Correlation energy for the He atom, calculated
both without (diamonds) and with (crosses) the correction 
term, using the PBE correlation functional.  The solid line is the exact
result, calculated by scaling the PBE correlation functional.}
\end{figure}
Finally, we discuss our results for the two-electron ion series,
simply to demonstrate that there is nothing special about Hooke's
atom which makes these techniques accurate.  We may make bare estimates
of the adiabatic curves directly from energy data already in the
literature.  Thus by calculating $\lambda$ by Eq. (\ref{ldef}) using the
data
in Table I of Ref. \cite{HU97}, and assigning to each $\lambda$ the bare
$U\c$ value from the table, we find the corresponding adiabatic
connection curve integrates to  $E\c=$ -41.5 mH, as compared with an
exact value of -42.1 mH for He.  We can similarly use the $T\c$ data
in Eq. (\ref{EcfromTcg})
to get an estimate of -41.2 mH, so that the difference between these
two results is a good indicator of the error in both of them.
Again, we can calculate $T\s\l$, and find the largest error at 
$\lambda=0$, with a value of 2.916, as compared to the exact value
of 2.867.  

In Fig. 6, we plot PBE calculations of $E\c[\n_{1/\lambda}]$
for the He atom, both exactly and using the fake adiabatic
connection formula, both with and without the correction term.
We use the highly accurate densities of Umrigar and Gonze\cite{UG94}.
Unfortunately, not enough data points are available to make
smooth plots for this system.   Already at $Z=3$,
$\lambda=0.62$, so that results for non-integer values of
$Z$ (between $Z=2$ and $3$) are needed.
However, enough data is present to see the
clear correction the potential makes.  We note several
interesting features of this curve.  First, the dependence
on scaling is much less for He than for the $k=1/4$ Hooke's
atom.  Second, the PBE curve contains a minimum, 
so Eq. (\ref{TcfromEcg}) implies
$E\c+T\c$ becomes positive for this scaled density.  While
it has never been rigorously proven, every known case of $E\c+T\c$ 
is negative\cite{LP85,B97}.
Thus this is probably a (very slight)
limitation of PBE, which does not occur in PW91, and may be related to
the difference at $r_s=0$ in Fig. 7 of 
Ref. \cite{PBW96}.
Third, the potential correction even picks up this feature,
and still only has errors of a fraction of a millihartree.

\begin{figure}[htb]
\unitlength1cm
\begin{picture}(12,6.5)
\put(-5.2,4.0){\makebox(12,6.5){
\includegraphics{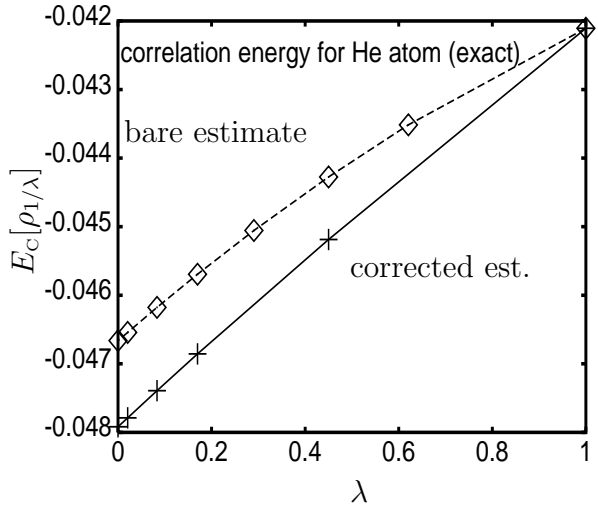}}}
\setbox6=\hbox{\large $E\c[\n_{1/\lambda}]$}
\put(0.8,3.85) {\makebox(0,0){\rotl 6}}
\put(5.1,0.1){\large $\lambda$}
\put(5.1,3.1){\large corrected est.}
\put(2.1,4.9){\large bare estimate}
\end{picture}
\label{ecHe}
\caption{Correlation energy for the He atom, calculated
both without (diamonds) and with (crosses) the correction 
term.}
\end{figure}
Figure 7 repeats Fig. 6, but for the exact case.  Here the
corrected curve is much straighter than for PBE, and has no
minimum.  The lines are drawn merely to aid the eye.
Note that the value of $E\c^{(2)}[\n]$, which is the high-density
limit of $E\c$, drops about 1 millihartree, due to the 
difference in shape between the He atom and a pure exponential
decay (bare estimate).   We find $E\c^{(2)}$= -47.9 millihartree,
in reasonable agreement with the value of 
Colonna and Savin\cite{CS99} 
(-47.5 from Table VI),
of Joubert and Srivastava\cite{JS98}
(-47.2 from $a_p$ in Table II),
and of Engel and Dreizler\cite{ED98}
(-48.2 from Table V).
The value -50.3, reported in Ref.
\cite{SPL99}, was evaluated on the PBE selfconsistent density.

\section{Conclusions and Implications}
\label{conc}

We have shown how, with accurate ground-state results as a function
of the external potential, 
an accurate adiabatic connection curve can be calculated for both Hooke's atom
at $k=1/4$ and the He atom.   We see no reason why similar results
could not
be obtained  for larger atoms, especially those for which the Kohn-Sham
potential has already been calculated\cite{ZMP94}.  There also
exist methods for isolating the correlation potential\cite{FUG96}.

In the event that the exchange contribution cannot be easily isolated,
the scheme can still be applied, but now to the combined exchange-correlation
energy at each value of the external parameter.  The correction 
is now evaluated using the exchange-correlation potential.  On Hooke's
atom, this yields results almost as accurate as those described for
the correlation energy alone.  Important differences are in the regime
$\lambda\to 0$, as here the exchange contribution to the correction
blows up, since $\gamma=1/\lambda \to \infty$.
This effect causes noticeable errors
only for $\lambda < 0.2$ in $E\c[\n_{1/\lambda}]$, but not in $U\c$ or
$T\c$.  It also means that neither $U\c[\n]$ or $T\c[\n]$, as derived
from the resulting $E\c[\n\g]$ curve
via Eqs. (\ref{UcfromEcg}) and (\ref{TcfromEcg}), is exact at
$\lambda=1$.  However, these
differences are of the order of 0.1 millihartree.

Finally, we are currently investigating if this method can be used 
to calculate accurate adiabatic connection curves for
molecules\cite{EPB97}.

\section*{Acknowledgments}
This work was supported by an award from
Research Corporation and by the National Science Foundation under
grant No. CHE-9875091.  We thank John Perdew for his comments,
and Cyrus Umrigar for providing us with accurate densities for the two
electron ions.

\end{document}